\documentclass[fleqn,usenatbib]{mnras}

\usepackage{newtxtext,newtxmath}

\usepackage[T1]{fontenc}
\usepackage{ae,aecompl}

\usepackage{graphicx}	
\usepackage{amsmath}	
\usepackage{amssymb}	

\title[Braking index jumps in young pulsars]{Braking index jumps in young pulsars}

\author[J. E. Horvath]{J.E. Horvath,$^{1}$\thanks{E-mail: foton@iag.usp.br} 
\\
$^{1}$Universidade de S\~ao Paulo, Department of Astronomy IAG-USP\\ R. do Mat\~ao 1226, 05508-090, Cidade Universit\'aria, S\~ao Paulo SP, Brazil
}

\date{Accepted XXX. Received YYY; in original form ZZZ}

\pubyear{2019}

\begin{document}
\label{firstpage}
\pagerange{\pageref{firstpage}--\pageref{lastpage}}
\maketitle

\begin{abstract}
The departure of all measured pulsar braking indexes from the ``canonical'' dipole value 3 has been attributed to several causes in the past. Careful monitoring of the Crab pulsar has revealed
permanent changes in the spin-down rate which are most likely the accumulation of small jumps in the angle $\alpha$ between the magnetic and spin axis. Recently, a large permanent change in the
braking index of the in the ``Crab twin'' pulsar B0540-69 has been reported, and an analogous phenomenon seen in the high-field pulsar PSR 1846-0258 has been seen following a glitch, while
another similar event (in PSR J119-6127) needs to be confirmed. We argue in this work that a common physical origin of all these observations can be attributed to the counter-alignment of the
axis without serious violations of the observed features and with very modest inferred values of the hypothesized jump in the $\alpha$ angle. In addition, detected increases of the X-ray luminosities after the events are an additional ingredient for this interpretation. We argue that a component of a time-dependent torque has been identified, being an important ingredient
towards a full solution of observed pulsar timing behavior which is in search of a consistent modeling.
\end{abstract}




\section{Introduction}

The timing of pulsars has revealed, among other phenomena, sudden discontinuities (called {\it glitches}) in the pulsar frequency $\Omega \; and \;$ spin-down rate $\dot \Omega$, which are
particularly conspicuous in young objects. Several of these events have been observed from the Crab, Vela and other pulsars. Because of long recovery timescales observed, it is now generally
agreed that they are the result of a (complex) interplay between the superfluid component(s) and the ``normal'' components of the star. Even though a variety of models have been proposed, the
most accepted have searched an explanation in terms of sudden decoupling of vortices in the superfluid component (\cite{AI, PinesAlpar}). Earlier models invoked cracking of the stellar crust
(\cite{GB}), and despite problems related to its viability, survived in a different form in later works (\cite{Mal}). The model in which there is a vortex creep in steady state, suddenly ended
by a critical condition is flexible enough to accommodate a large body of observations, but evidence coming from very timely and detailed observations of glitches in the Crab pulsar, B0540-69
and PSR J1846-0258 seem now to indicate that new physical inputs are required to explain the data, at least concerning the recovery behavior.

As stated above, the observations of post-glitch recovery in some of these objects have revealed {\it permanent} changes in the spin-down frequency and its rate. As a consequence, the pulsars
rotate differently after the glitches, indicating a change in the dynamics, i.e. some change in the braking torque. The ``permanent'' changes related to angle variations will be the object of
our discussion.

Since its very discovery (see \citet{Dick} and references therein), the glitch phenomenon was studied to deepen the understanding of the pulsar dynamics. Presently the number of glitches has
grown very substantially and the number of pulsars suffering glitches is at least 180 (\cite{Dick}). The hypothesis that {\it all} pulsars have glitches is still tenable given the time coverage
of the whole pulsar sample, $\geq 10$ times larger than the glitching sample and growing. While larger glitches have been observed in older pulsars, the latter seem to be sporadic, while young
objects tend to be frequent glitchers. Moreover, smaller but more frequent glitches are observed from Crab-aged pulsars compared to the elder objects with ages $\geq 10^{4} -10^{5} \, yr$ such
as the Vela pulsar and similar objects (\cite{Espi}).

The group of pulsars showing a permanent change in the frequency derivative is small, but important enough to ask the question of the causes of such a difference, its consequences for the
braking index (which have been observed for the first time) and the general trend of young pulsars dynamics. It includes the Crab pulsar, PSR 1846-0258 and PSR J1119-6127, while other similar
phenomena have been reported for B0540-69 (the ``Crab Twin'') and other intermittent pulsars. The case of PSR J1846-0258 is particularly interesting because it is a ''high-field'' pulsar,
sitting in the ${P-{\dot{P}}}$ diagram right below the magnetar group. We shall see that these glitch events could reveal important features of the latter objects indirectly. The subject of our
discussion here is to argue that all these permanent changes and their associated changes in the measured braking index (confirmed in the Crab, PSR J1846-0258 and yet pending confirmation in a
PSR J1119-6127 glitch) are quite naturally explained in terms of the discrete counter-alignment of the magnetic and spin axes.

It is important to note that the hypothesis of the misalignment of the axes has been considered in the past (\cite{DG, MG, Curt, Macy, LM, BR, AH, JK}), and some physical reasons for it to occur
discussed. One of the simplest physical realizations, namely through plate tectonics, also has a quite extensive history (\cite{Mal}). One interesting argument supporting a plate tectonics
physical origin of it has been presented by \cite{LE97}, showing that the general glitch pattern is akin to earthquake behavior. The possible association of discrete jumps with a particular
mechanism of glitch is not restricted to a ``starquake'' driven by crust plate tectonics, but it is difficult to picture for others, for example, a thermally activated mechanism \cite{BR}. The
positive detection of an evolution of the braking index is the main reason to revisit and re-discuss this scenario.

\section{General framework}

As is well-known \citet{manchester}, the torque equation of the pulsar is simply

\begin{equation}
I_{tot} \dot \Omega \; = \; \tau_{ext}  .
\end{equation}

where $I_{tot}$ is the sum of the components which are rigidly coupled to the crust of the star and $\tau_{ext}$ has to be calculated by integrating the Poynting vector over a surrounding
surface, adopting in general the form $\tau_{ext} \; = \; -K \Omega_{c}^{n}$. The ``pure'' vacuum dipole expression is just $K \, = \, {\frac {2}{3 \, c^{3}}} \, \left\vert m \right\vert^{2} \,
\sin^{2} \alpha$, where $\alpha$ is the angle between $m$ and $\Omega_{c}$ ,
$\left\vert m \right\vert \, = \, B_{o} \, R^{3}$ and $B_{o}$ , $R$ are the magnetic field and the radius of the star respectively. Even when the (compelling) presence of charges is considered
in the magnetosphere, the dependence of the torque remains similar, and therefore a pulsar braking by dipole radiation only, with a prefactor $K$ independent of time, is governed by $n = 3$.
Within this well-known framework, the accuracy of the dipole-braking picture can be further checked observationally by expressing the braking index as $n_{obs} = {\frac{\ddot{\nu}{\nu}}{\dot
\nu}}$, where we have labeled with the suffix ``obs'' the observed braking index to distinguish it from the the theoretical value $n = 3$, in which none of the parameters depend on time.

The difference between the actual power in the theoretical torque equation (1) and the {\it operational} definition in terms of observable quantities can not be overstated (\cite{AH}). Of
course, even if the electromagnetic dipole is the only component, (i.e. without wind or other additive components), there is still the possibility of ``non-standard'' effects if one of more of
the structural quantities in $K$ depend on time, affecting the value of $n_{obs}$. Quite generally, the latter possibility can be written as \cite{AH,Lyne45}

\begin{equation}
n_{obs}= n \, + \, {{{\ddot \nu} \nu} \over{{\dot \nu}^{2}}} \, = \, n \, + {\biggl( {-{{\dot I} \over{I}} + 2 {{\dot \alpha}\over {\tan \alpha}} +2 {{\dot m} \over{m}}}  \biggr)}
\end{equation}

We can now specify for the particular case of angle variation only (second term inside the parentheses), making 
${\dot I} = {\dot m} = 0$, 
and using the notation of \citet{AH}, eq.(2) can be inverted to relate the angle jump quantities to the
observable quantities

\begin{equation}
{{\Delta\alpha} \over {\Delta t}}{1 \over{\tan \alpha}} \, = \, {{n_{obs} - 3} \over 2} {{\dot
\Omega} \over \Omega}
\end{equation}

where we have written $\Delta\alpha$ to indicate a (small) discrete jump in the angle across a glitch. We stress again that this expression assumes that the only non-standard effect leading to $n_{obs} \leq 3$ is just the
evolution of the angle $\alpha$, but makes no further assumptions about its physical cause. It is useful for an analysis of actual events in which sudden variations of the observed braking index
were detected, and its ``continuous'' version already used (\cite{AH}) to address the limiting situation of successive very small jumps (as inferred, for example, in the case of the Crab pulsar) is straightforward. This is the simple basic mathematical picture from which the behavior of the angle with time consistent with observations can be inferred.

\section{Discrete jumps in observed pulsars}

\noindent
\subsection{B0531+21 (Crab)}

The Crab pulsar is by far the best known example of a young radio pulsar experiencing glitches at a high rate. A recent summary of the timing history of the Crab has been presented by Lyne et
al. (2015). According to these authors, the accumulated glitch activity (some of which record a permanent change in $\dot \Omega$ across the glitch) leads to a $\sim 6\%$ overall decrease of the
spin-down rate, lowering the mean braking index to 2.34 for the last 45 yr from the quoted steady value of 2.51 (\cite{Maggie, Lyne45}). The case of the Crab has been modeled in \cite{AH} using a
quasi-continuous expression which adds up small changes in the spin-down, further attributed to a variation of the angle $\alpha$ driven by misalignment (actually the same interpretation favored
now by \cite{Lyne45}). Using a time average and interpreting the time interval between glitches $\Delta t \sim 1 yr$ as the relevant timescale, we estimate from eq.(3)

\begin{equation}
{{\Delta\alpha} \over {\Delta t}}{1 \over{\tan \alpha}} \, = 9 \times 10^{-5} rad \, yr^{-1}
\end{equation}

The emerging picture for the Crab braking behavior is that the accumulating effect of small changes over $\sim$ decades timescale led to the decrease of the observed braking index $n_{obs}$. The
counter-alignment of both axes is a working basic hypothesis, which explains this behavior and yields a reduced braking index. As pointed out in \cite{AH}, the angle seems to increase not only
across the glitches, but also between them. As a consequence of the steady decrease in the braking index, \cite{Lyne45} pointed out that within a few thousand years this behavior would drive the
Crab towards the magnetar region of the $P-{\dot P}$ diagram, but without implying a huge magnetic field. The latter kind of behavior was suggested from the analysis of models including a wind
component \cite{SB}, although it actually holds for all models in which an evolving braking index with the right properties ensue (see, for example, \cite{Ze}). This ``magnetar'' issue is of
course controversial, since it is possible to argue that a number of features observed in SGR/AXPs do need a very high magnetic field indeed. Nevertheless, it is safe to sate that there is
enough empirical evidence for an heterodox set of trajectories in the $P-{\dot P}$ plane just as a consequence of varying braking indexes.

\bigskip
\noindent
\subsection{PSR 1846-0258}

This high-field pulsar was observed to glitch in 2006 (\citet{Gav}), emitting several bursts, and was later monitored by \cite{Arch1} for 7 years. The pulsar featured a jump in $n_{obs}$ across
the 2006 glitch amounting to $\Delta n_{obs} = 0.46 \pm 0.03$ (\cite{Liv1, Arch1}), with a large statistical significance. Within the same picture as above, we can apply the simple formulae to
obtain

\begin{equation}
{{\dot \alpha} \over{\tan \alpha}}  \, = 8.74 \times 10^{-11} rad \, s^{-1}
\end{equation}

\noindent
or equivalently

\begin{equation}
\Delta \alpha = 2.76 \times 10^{-3} {\biggl( {\Delta t \over{1 \, yr}} \biggr)} \tan \alpha \, rad .
\end{equation}

Thus we see that the large jump in $n_{obs}$ also requires a very small jump in the angle $\alpha$, which nevertheless needs to be an order of magnitude larger than the ones inferred for the
young Crab. The present caveats about the value of the angle itself, reflected in the $\tan \alpha$ factor would need more observational input to be solved. In particular, a detailed modeling of
the pulse in several bands is in order to determine, or at least put limits, on the value of the angle itself.

\bigskip
\noindent
\subsection{PSR J1119-6127}

The last concrete example of a discrete behavior in the braking index was reported by Antonopoulou et al. (2015) in this high-field pulsar. The 2007 glitch was accompanied by a decreasing change
$\Delta n_{obs}$ of around $15\%$. Even though the object is quite similar to PSR 1846-0258, the after-glitch recovery was quite different, as these authors discuss. This delicate issue is, of
course, ultimately related to the glitch mechanism itself, which is beyond the scope of this work. Looking to this jump alone we find that

\begin{equation}
{{\dot \alpha} \over{\tan \alpha}}  \, = 1.5 \times 10^{-12} rad \, s^{-1}
\end{equation}

A feature of this pulsar which adds some information to the estimates is that the angle $\alpha$ is {\it not} close to $90^o$, but rather believed to be bounded to $\leq 30^o$ (\cite{WJE}).
Therefore we can be confident that counter-alignment has not occurred and the $\tan \alpha$ is not $ \geq 1$. We see that the numbers are in line with the ones estimated for the other cases,
pointing toward very small jumps in $\alpha$ consistently.

\bigskip
\noindent
\subsection{ The case of B0540-69 (``Crab twin'') and other intermittent pulsars}

The discovery of a group of pulsars which alternate between two different spin-down states (see \cite{Lyne} for a comprehensive account) has been tentatively related to a change in the
spin-magnetic axis angle. This phenomenon shares the feature of a large spin-down change with the above group, but is different in a few important aspects. A prime example is the ``Crab twin''
B0540-69 in which a significant, sudden change in ${\dot \Omega}$ was observed on Dec 2011 (\cite{Marsh1}). Strictly speaking, this was not a common glitch since the spin-down frequency did not
change during the state transition. Nevertheless, angle changes were among the discussed causes (for example, in \cite{Marsh2}) for this remarkable event. The pre-event value $n_{obs, pre}=
2.129 \pm 0.012$ (\cite{Ferd}) became a very low $n_{obs,post} = 0.031 \pm 0.013$ (\cite{Marsh2}. This is an unprecedented behavior for any pulsar, and could be considered a ``jumbo'' version of
the events of the Crab described above. Assuming a single angle jump (therefore dropping the average symbol) and inserting the measured values of $\Omega$ and ${\dot \Omega}$, eq.(3) yields at
face value

\begin{equation}
\Delta \alpha = 3 \times 10^{-4} {\biggl( {\Delta t \over{1 \, yr}} \biggr)} \tan \alpha \, rad .
\end{equation}

Therefore, unless the spin and magnetic axes are already almost orthogonal, the actual required angle jump is actually minuscule. The timescale in parenthesis scaled to $1 \, yr$ can further
suppress the required value by several orders of magnitude if identified now with the time required to complete the transition, which in some cases lasted $\sim days$ at most. The difference
with the Crab case in which a series of events have been treated as a quasi-continuum is evident. Nevertheless, the bottom line here is that a single, albeit {\it very small} change in the angle
can account for the observed $\Delta n_{obs}$, therefore it may not be surprising that the average pulse itself did not show any measurable change.The absence of a change in the frequency itself
may put this event in line with the so-called ``intermittent'' pulsars (\cite{Lyne}), as pointed out by Marshall et al. (2015). While models with an extra torque look a promising way to account
for these kind of phenomenon, we believe that the important observation is the lack of a significant increase in the steady X-ray flux after the transient (\citet{Liv1}). The ``jump'' in the
angle $\alpha$ faces a similar problem, since it should cause an opening of the field lines ultimately related to an increase of $L_{X}$. This issue is still open, but we are reluctant to
include intermittent pulsars as examples of angle misalignment because of the latter issue. On the other hand, it is recognized that a magnetospheric rearrangement preserving $\Omega$ and
$L_{X}$ simultaneously, while changing ${\dot \Omega}$ is not easy to envisage.

\section{Conclusions}

We have discussed in this work the interpretation of discrete jumps in the angle between magnetic dipole and spin axis as an ingredient to understand the braking indexes of (young) pulsars. The
most extreme case of the angle jump is associated to PSR 1846-0258, an order of magnitude higher that the mini-jumps needed to explain the Crab pulsar behavior. However, even in this most extreme
case the required jump is still small enough, and a noticeable change in the average pulse profile is not necessarily expected. Only if the angle itself is $\sim \pi/2$, making the $\tan \alpha$ a very large number, the small jump ${\dot \alpha}$ would produce dramatic effects on the pulse, but the former is unlikely to be the case.This is our main conclusion about the angle jumps: they constitute a very small, accumulative effects which induce secular, progressive small changes in the pulse profile at most.

We suggest that the common features in this sample of pulsars are an important, unifying hint of a dynamical behavior which is not included in the ``standard'' view of pulsar spin-down. It is
further indicated that pulsars older by a factor of $\sim 10$ than the group discussed above could hold the clue of the limiting age beyond which the misalignment stops (\citet{AH, Linketal}).
Moreover, there is even a hint of a anti-correlation with the characteristic age (see Table 1) in this group of objects, and although it should not be taken very seriously until the sample can be
enlarged, it is suggestive of a {\it decreasing} trend in the magnitude of the angle jump with age, i.e. an evolution of the misalignment by requiring smaller jumps along the pulsar's life until
the complete misalignment is reached around $\sim 10^{4} yr$. This is why to confirm that Vela or similar age pulsars do or do not display permanent changes after glitching is so important for
the dynamics of the young pulsars, a negative answer would point towards ``maturity'' in the sense of completed misalignment of the spin and magnetic axes. We have learned from these events that
braking is indeed variable in (all?) young pulsars, and that while the rate of glitches seem to decrease with age, they may involve progressively smaller jumps in the $\alpha$ angle.

Alternatively to the angle-growth, the sudden decoupling of some internal component leading to ${\dot I}$ (first term between parentheses of eq.2, \cite{Heitor}) has been put forward as a
contender, but has been questioned before (\citet{LE97, Gordon}) because within the latter hypothesis the torque equation does not allow to interpret the spin-down behavior seen after the
glitches easily. On the other hand, the growth of the magnetic field represented by the third term between parentheses of eq.(2) has been discussed in the literature on a physically sound basis
(see, for instance, \cite{Eksi}). However, it is important to stress that the X-ray luminosity have been confirmed to {\it increase} after the events of PSR 1846-0258 and PSR J119-6127, as
expected from the angle-jump model loosely predicting $L_{X} \propto \sin^{2} \alpha$ (\cite{SB}), a feature which is difficult to explain by jumps in the moment of inertia, or even by a sudden
increase in the magnetic field growth in which $L_{X} \propto B \Delta B$ only, which would call for substantial value of the former quantity. In this sense, it can be stated that the X-ray
luminosity observations give support to the angle-growth hypothesis, favoring it over its contenders.

The knowledge of the overall pulsar behavior has grown a lot in the last 20 years or so. The database comprise now more than 2000 thousand objects, and their timing monitoring identified
hundreds of glitches, allowing a general picture based on statistics (\cite{Espi}). The glitching behavior varies widely from object to object, but it is fair to state that the incompleteness of
the basic dipole-braking model is clear and in search of complementary work. While it is somewhat dangerous to generalize, we believe that one additional ingredient, the variation of the
spin-magnetic angle, is a leading contender among the non-standard effects. Additional components of the torque, i.e. winds, not addressed here may also contribute to the braking. Whatever the
final verdict is, it is time to go beyond the zeroth-order dipole-braking picture in this problem at once.

The varying angle is another phenomenon that would contribute to the so-called timing noise (i.e. changes in the timing residuals and rotation frequency, \cite{Ly}). It is to be regarded as inducing a (small) torque variation, a non-standard 
behavior in the canonical models. Sudden changes in the magnetospheric states have also been reported (\cite{Kramer}), with a switching of the braking index between two different values. It is unclear if any relationship between angle jumps and 
magnetospheric current switches exists, the latter seem to require a large modification but happens to be reversible. 
In any case, at least two phenomena that can affect the braking index in a discrete fashion are now known. We can now 
ask the question whether all timing noise can be generated by these discrete changes. Lyne (2013) answers positively to this hypothesis considering magnetospheric switching only, and we may add that young pulsars timing noise may be a consequence of angle variations, even very small undetected ones. A physical picture of the cause of angle variation (i.e. plate tectonics, \cite{LE97, Mal}) can lead to a natural expectation for the latter expectation.

Finally we would like to remark that the non-standard trajectories bending upwards in the $P-{\dot{P}}$ plane do exist and would put a number of $\sim 10^{3} yr$ objects in the magnetar region
within a few millennia (see \cite{Crist} and references therein). The time-dependence of the torque would also call for a significant correction to all characteristic ages, since the latter
assume {\it constant} values of the braking index within the canonical approach.The ages should now require the determination of additional quantities, as discussed in Allen and Horvath (1997)
(see also \cite{Crist} for a recent extended account), and could be an important factor to understand the well-known mismatches between the pulsar and supernovae ages.

\begin{table}
	\centering 	\caption{The characteristic ages and angle jump values of the three cases addressed in the text} 	\label{tab:example_table} 	
	\begin{tabular}{lccr} 
	\hline 	    Pulsar Name & Characteristic Age (yr)&  ${{\dot \alpha} \over{\tan \alpha}}\, (rad \, yr^{-1})$\\ 		
	\hline 		PSR1846-0258 & $\sim$ 800 & $2.76 \times 10^{-3}$\\ 		
	                B0531+21 & 1240 & $9 \times 10^{-5}$  \\ 		
                     PSR J119-6127 & 1600-1900 & $4.7 \times 10^{-5}$ \\ 		
	\hline 	\end{tabular}
\end{table}

\section*{Acknowledgements}

The author wish to acknowledge the financial support of the Fapesp Agency (S\~ao Paulo) through the grant 13/26258-4 and the CNPq (Federal Government) for the award of a Research Fellowship. We are indebted to an anonymous referee which has contributed to shape this paper with constructive criticisms and remarks, including the rising of the issue of the timing noise in the last Section.





\bsp	
\label{lastpage}
\end{document}